\documentstyle[11pt,moriond,epsfig]{article}

\def\be{\begin{equation}}
\def\ee{\end{equation}}
\def\bea{\begin{eqnarray}}
\def\eea{\end{eqnarray}}

\begin{document}
\vspace*{4cm}
\title{SUMMARY OF THE EXPERIMENTAL  
       PART\\ 
       OF THE XXXIVth RENCONTRE DE MORIOND.} 

\author{ M.W. Krasny }

\address{Balliol College and Particle and Nuclear Physics 
Laboratory, Keble Road,\\
Oxford OX1 3RH, England \\
and \\
L.P.N.H.E, Universities Paris VI et VII, 4 pl. Jussieu, \\
75252 Paris, France}

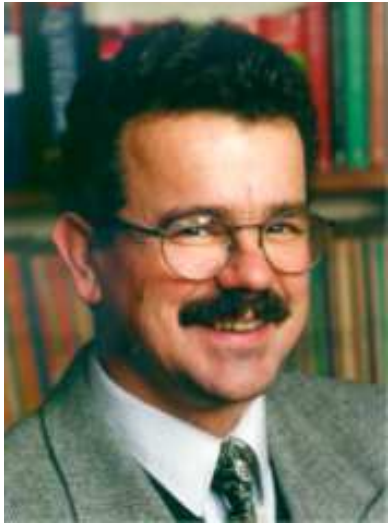
\maketitle\abstracts{
I summarise the experimental results presented during the hadronic session
of the XXXIV Rencontre de Moriond.
\footnote{Summary talk at Moriond, 
Les Arcs, 20-27 March, 1999. }
 }
\newpage 

\section{Introduction}

The XXXIV Rencontre de Moriond, the last in the current millennium,
will be remembered  
by all of us, as the time of 
joy  of sharing the curiosity about the processes of the subatomic world.
The invariantly open character  of Moriond meetings, in which 
mostly young promising physicists
are given a chance of not only 
presenting results of their collaborations
but also exposing  their own ideas and views in an
informal atmosphere, reveals unique spirit of a common cause 
which is often hidden during the preceding years of tedious 
and competitive work. 
On behalf of all participants I offer my thanks 
to the organisers of Recontres de Moriond: to their  {\it chef} 
Jean Tr\^an Thanh V\^an and to his {\it \'equipe}.
May these successful meetings 
carry on in the coming millennium.  

The experimental part of the session 
devoted to ''QCD and High Energy Hadronic Interactions" 
was spanned between three poles of activities. 
QCD, the theory of interactions of 
sub-hadronic coloured quanta
and one of the pillars of the standard model, describes 
an impressively large variety of phenomena providing 
precise quantitative predictions. A large fraction 
of  measurements presented
at the conference were motivated by and optimised for the 
most precise quantitative tests of the 
perturbative QCD predictions. 
Measurements which map partonic distributions of extended 
hadronic objects and analyses which 
tried to provide comprehensible classification  of non-perturbative
effects represented the second pole of activities.
They both establish  important relationships between 
processes involving  colliding  particles of variable 
types and energies.  
 The third class  of measurements included those  
driven by pure curiosity.
 Most of them lay beyond the regime of applicability of perturbative 
QCD. 
With some regret,
which was shared by several conference participants,
I note  that the fraction 
of Moriond-presented-results which belong to this type 
of activity keeps decreasing. Searches for the  origins
of several puzzling strong interaction  
phenomena are very often replaced by 
the ''searches" of the most appropriate Monte-Carlo
parameters 
to  describe, rather than explain, these phenomena.

A large majority of results presented during the conference are 
summarised here, but
with some embarrassment, several of the the contributions
had to be left out. The results on heavy flavour decays were 
summarised in 
Guido Martinelli's concluding theory talk and are not included here. 
Contributions devoted to physics at  
future machines were omitted deliberately.
Predefined scenarios of what the physics at the next generation
of colliders is going to be is of non-questionable importance at the 
time of designing the detectors. But 
searches for {\it anticipated signatures} of new physics
will have to be complemented, as soon as new colliders become operational,
by searches for unexpected phenomena driven by 
experimentalist's curiosity and unbounded by 
{\it the Beyond the Standard Model
Theory-Guides}. I would like to express 
the hope of several of the conference participants
that the results of such searches will add a lot of excitement to the 
Rencontres in the first decade of the next millennium.

\section{Standard Model Parameters}

Measurements of the standard model parameters were extensively 
discussed during the electro-weak session
preceding the hadronic one.
The most important results were summarised at the 
hadronic session. 
S. Choi \cite{Choi} and T. Dorigo \cite{Dorigo} presented 
the CDF and D$\O$ measurements of the masses of the top quark and 
of the W boson
while T. Saeki \cite{Saeki} discussed W mass measurements 
by the LEP experiments. The  results of the standard model fits
to the data were presented by G. Della Rica \cite{RICA}.

The top quark is an outstanding member of the quark family. 
To a certain extent it resembles more a lepton than 
its family fellows. In its short life,
the top
quark is not exposed to the colour confining forces
which affect all other family members.
Thus, its  mass is measured with 
a precision which 
can not be reached for lighter quarks. 
The reported average CDF/D$\O$ mass is: 
$M_{t}= 174.3 \pm 3.2 (stat) \pm 4.0 (syst)$ GeV.
Once the value of the top mass is fixed its 
production cross-section is quite precisely
predicted by the standard model. The reported D$\O$ and CDF
cross-section 
measurements agree with the standard model expectations, 
extending   
an important test  of the universality of quark interactions to 
five orders of magnitudes of their mass span. 

The W mass measurements are summarised in Table \ref{wmass}. 
The consistency of all  direct and indirect 
measurements  is impressive,  given the high 
precision of the  reported results and allowing the LEPII
measurements to be  potentially affected by the colour
reconnection effects and/or by correlated fragmentation of quarks
originating from different W bosons. A comprehensive overview of 
searches for the above effects was given by W. Kittel \cite{Kittel}.
The results of the four LEP experiments, reported by F. Martin
\cite{Martin1},
are rather confusing. DELPHI results favour Bose-Einstein
correlations between pions originating from  different W bosons
at 2.4$\sigma$ level while ALEPH results, obtained
using different methods and variables, disfavour   
such correlations at the 2.7$\sigma$ level. Moreover,  only the DELPHI
collaboration observes a 2$\sigma$ difference of the  
average charged particle multiplicities 
of the hadronic decays of W bosons in 
$ e^{+}e^{-} \rightarrow  q \bar q q \bar q$ and 
$ e^{+}e^{-} \rightarrow  q \bar q l \nu $ reactions. 

\begin{table}[t]
\caption{Direct and indirect measurements of the W mass. \label{wmass}}
\vspace{0.4cm}
\begin{center}
\begin{tabular}{|c|c|}

\hline
&   \\
Experiment & $M_{W}$  (GeV) 
\\ \hline

&   \\
$p \bar p$-colliders (direct) & 80.448 $\pm$ 0.062 
\\ \hline

&   \\
LEP2 (direct)& 80.370 $\pm$ 0.063 
\\ \hline

&   \\
NuTEV/CCFR (indirect)& 80.25 $\pm$ 0.11 
\\ \hline

&   \\
LEP1/SLD (indirect)& 80.326 $\pm$ 0.037 
\\ \hline

\end{tabular}
\end{center}
\end{table}

G. Della-Rica \cite{RICA} presented  
an update of the standard model fits to the world data. With 
respect to the  1998  summer conferences 
two changes have to be mentioned: 
theoretical improvements in ZFITTER 5.20  package, 
which was used in the fits,
and an increased precision of the $M_W$ measurements.
The fit parameters are shown in Table \ref{fits}. 
What is particularly noteworthy is not only 
the remarkable success of the standard model in explaining
the consistency among a large number of, {\it a priori}
uncorrelated,  measurements but, in addition, that a single, 
spin-less particle, with the mass of 
$M_H = 71^{+75}_{-42}$ GeV, predicted by the simplest mechanism of 
spontaneous electro-weak
symmetry breaking, is both necessary and sufficient
to achieve such a degree of consistency.   
Whether or not this 
particular symmetry breaking mechanism
is the one chosen by Nature remains, however, to be 
experimentally demonstrated - there is still  room 
for surprises here.
It remains to be added that 
further improvements 
of the precision of the fit results 
are expected to come from the ongoing measurements of the 
cross-section for $e^{+}e^{-}$ annihilation at low energies.
These measurements should  
improve our knowledge of the electro-magnetic coupling constant 
$\alpha_{EM}$.
The preliminary 
results from BES - a newcomer at Rencontres -
and prospects for future improvements were discussed by D. Paluselli
\cite{Paluselli}.

Y. Gao \cite{Gao} presented a new measurement of $\alpha_s$ at LEPII.
The combined LEP result derived from the data taken at 189 GeV,
$\alpha_s(189$ GeV$)= 0.1084 \pm 0.0040$ and corresponding to 
$\alpha_s(M_Z)= 0.121 \pm 0.005$, agrees  with the 
LEPI result. The energy scale dependence of $\alpha_s$ 
determined using both the  LEPI and LEPII data  was 
shown to be consistent with the perturbative QCD predictions.

In QCD the strength of the coupling of quarks 
to gluons is independent of the quark flavour. The data collected
at LEP and SLC and showed  at this conference by 
S. Cabrera-Urb\'an \cite{cabrera} confirm 
this universality with a precision 
better than 3 $\%$.

\section{Fundamental Standard Model Processes at LEP, SLC  and HERA}

The measurement of the cross-section for the production of a pair
of W  bosons in $e^+e^-$ annihilation
provides a very important test of the standard
model. It verifies the  
non-abelian character of the theory and 
tests its predicted pattern of self-couplings
of the carriers of electro-weak 
forces. 
The expected centre-of-mass energy dependence of the 
cross-section reflects an interplay between the 
t-channel, $\nu$-exchange diagram and the triple boson 
$ZWW$ and $\gamma WW$ diagrams. 
The combined LEP cross-section 
at 183 GeV of 15.83 $\pm$ 0.36 pb,
reported by Y. Uchida \cite{Uchida},
agrees perfectly with the standard model predictions. 
The combined preliminary LEP cross-section at 189 GeV of
 16.65 $\pm$ 0.33 pb
is somewhat lower than the predicted value - although no disagreement
can be claimed. The L3 preliminary  measurements of the 
Z-pair  production cross-sections at 183 and 189 GeV are in a good
agreement with the theoretical predictions 
within large statistical errors.   

\begin{table}[t]
\caption{Standard Model Fits. \label{fits}}
\vspace{0.4cm}
\begin{center}
\begin{tabular}{|c|c|c|}

\hline
& &   \\
SM parameter & fitted value & direct measurement  

\\ \hline

\hline
& &  \\
$M_{t}$  & 171.7 $\pm$ 4.9  GeV & 174.3 $\pm$ 5.1 GeV
\\ \hline

& &  \\\
$M_{H}$ & $71~^{+75}_{-42}$ GeV &  ???  
\\ \hline

& &  \\\
$sin^2 \theta ^{lept}_{eff}$ & 0.23154 $\pm$ 0.00018 & 0.23157 $\pm$ 0.00018
\\ \hline

& &  \\\
$M_W$ & 80.378 $\pm$ 0.0024 GeV & 80.410 $\pm$ 0.044 GeV
\\ \hline

\end{tabular}
\end{center}
\end{table}

C. Niebuhr \cite{Niebuhr} 
presented the results of  H1 and ZEUS experiments
on large $Q^2$ ($Q^2 \geq M_W^2, M_Z^2$) 
electron-proton and positron-proton scattering. After three
years of positron runs at HERA and following the 
necessary upgrade of the machine
vacuum system to store large electron currents    
the HERA detectors registered  in 1998 and 1999     
collisions  of electrons with  protons. The  high  
$Q^2$ $e^{-}p$ cross-section is expected to be larger
than that for the $e^{+}p$ collisions, both for the Neutral Current
interactions (due to the 
positive sign of the $Z^o/\gamma$ interference term)
and for the Charged Current interactions (dominated at high $x_{Bj}$
by $u$ rather than by $d$ quarks).
Even if HERA results have little impact on precision 
tests of the standard model, the measured $e^{+}/e^{-}$ 
cross-section ratios 
cross-check  the  magnitude of the t-channel  
 $Z^o/\gamma$ interference term   
and provide the first glimpse at the flavour 
structure of the proton in the kinematical region 
which is beyond the reach 
of the large statistics fixed target DIS experiments.

The SLD results on  forward-backward strange quark asymmetry 
at the $Z^o$ peak,
discussed by H. Staengle \cite{Staengele},
highlighted the merits of beam polarisation 
in improving the accuracy of this  measurement. The reported  
value of $A_s = 0.82~ \pm ~0.13$ is in agreement with the 
standard model prediction  and  is the most precise 
measurement of this quantity - in spite 
of the fact that the number  of events used
in this measurement was a factor 10 smaller than that of LEP 
experiments. 

Another interesting  SLD result reported 
at this conference by D. Dong \cite{Dong} was  
the measurement 
of the b-quark fragmentation spectra.
The reported average scaled momentum of the   
B-hadron of $<x_B> = 0.713 \pm 0.005 (stat) \pm 0.007 (syst)
\pm 0.002 (model)$
is the most 
precise direct measurement of this quantity.
It illustrates the  remarkable performance of 
the SLD CCD-Pixel Vertex Detector.

\section{Searches}

\subsection{Exotic states}

In view of the negative results 
of searches for the  ''manifestations of the 5D graviton in 4D",
F. Close \cite{close} devoted a major part of his talk 
to the role of pomerons in the formation of glueballs.
Confronted with a  ''Cinderella - task"  of  
filtering glueball candidates out of the  
known $q \bar q$ states, 
he argued that a sensible method of 
glueball hunting was to look at the initial state dependence
of the observed resonance pattern.  
In his view, by selecting glue-enriched
collisions, in particular 
those in which gluons 
have kinematicaly-enhanced coalescence probability, and 
by looking at the disappearance of a resonant signal for 
the photon-induced reactions the glueball spectrum enigma
might be deciphered.
A. Kirk \cite{Kirk} discussed   
the resolving power of the {\it coalescence filter} 
defined, for the reaction
$h_1 h_2 \rightarrow h_1 h_2 + R$, as    
the  transverse momentum difference of the 
outgoing $h_1$ and $h_2$ hadrons and interpreted by  Close  
as  the relative transverse momentum of 
the two recoiling pomerons.
He presented  the analysis of the
data from the WA102 experiment
and demonstrated  that by selecting  events with small 
relative transverse momentum of the two pomerons 
the $0^{++}$ and $2^{++}$ glueball candidates
can be filtered out   
from the known $q \bar q$ resonances.
  
Glueball searches in $\gamma \gamma$ collisions at LEP were
reported by D. Della Volpe
\cite{Volpe}. 
The mass spectra were analysed using the  {\it stickiness}, $S$, variable
measuring the relative coupling of a resonance to a pair 
of gluons and to a pair of photons. 
The disappearance of the $\zeta(2230)$
resonance both in the LEPI and LEPII $\gamma \gamma$ data
quantified by the average   
 LEP stickiness value of   
$S_{\zeta} \geq 68$ at 95 $\%$ confidence level, establishes 
$\zeta(2230)$ to be  a likely glueball candidate.

Are there 
colour singlet bound states of coloured constituents
other than mesons, baryons and glueballs?
QCD does not forbid the existence of hadrons composed 
of six quarks. As R. Ben-David \cite{Ben} argued, including strangeness
in constructing the  colour singlet $H^o=uuddss$ state may increase
the binding energy due to the colour-hyperfine interactions.
The reported searches for  such a state in the data of KTeV
collaboration ruled out the remaining mass window for a long 
lived $H^o$ as proposed by Donoghue et al. \cite{dummy}.
 
\subsection{Exotic strong interaction phenomena}
 
Traditionally the QCD-Rencontres coexist with Biology-Rencontres
and the common session gives a lot of joy both to biologists
and to physicists. 
Some of the ideas 
turned out, unexpectedly,
to be  common to both disciplines. S. Todorova-Nova
\cite{Todorova}
discussed possible experimental evidence of a (single) helix 
structure of the colour string proposed by Andersson et al. \cite{dummy}.

Modelling   
string fragmentation and multi-particle production processes
were implicitly present in  a large fraction  of results reported 
at this conference. A  silent assumption made
by  most of us, which is regretfully  
becoming a common, self-assuring consensus,   
is  that 
these processes are well controlled by the available Monte-Carlo 
generators.
Traditionally  
one of the basic questions of strong interaction
physics: how are hadrons produced from quarks and gluons?
is becoming routinely  
replaced by the question: how do we get rid of or minimise 
the hadronisation effects considered as  measurement noise
for jet-spectra, W-mass and several other measurements.
The LEP data on baryon production presented at this conference by 
R. Reinhardt
\cite{Reinhart}, showing   a 5$\sigma$ enhancement of proton production 
in gluon jets with respect to the quark jets
and a failure of existing generators to describe this data,
reminded us once more how limited our 
understanding of the 
particle production mechanism  is and ... why 
I included this result in the  
''exotic strong interaction phenomena" section of this repport.

\subsection{Dedicated searches }

Dedicated searches for  
the standard model Higgs and 
for new phenomena
predicted by various   
theoretical scenarios of 
beyond the standard model physics
were reported to give  negative results. 
The following particles (phenomena) have not (yet)
been found:

\begin{itemize} 
\item 
Higgs bosons (reported by I. Nakamura \cite{Makamura} for LEP
              experiments and J.Cassada \cite{Cassada} for FNAL
	      experiments);
\item
Supersymmetric particles appearing in MSSM, $R_p$ violating  MSSM,
GMBS, SUGRA and $R_p$ violating SUGRA (reported for the  LEP 
experiments by R. Alemany
\cite{Alemany}, for the HERA experiments by C. Niebuhr \cite{Niebuhr}and
for the FNAL experiments by R. Genik \cite{Genik}); 
\item
Technicolour, Topcolour and Flavour Universal Coloron (FUC) particles 
(reported by  J.Cassada \cite{Cassada});
\item 
Leptoquarks and 30 different Contact Interaction scenarios
(reported by C. Niebuhr \cite{Niebuhr});
\item
F. Close's 5D gravitons. 
\end{itemize}

The range of the above searches is indeed impressive. But what is 
the probability  that important new phenomena,
which can not be reduced to one of the above  scenarios, 
are overlooked? In my view such a probability is larger than one
might  naively expect.  
One of the reasons is that
in  {\it the rejection-limit-optimised} dedicated searches
a sizable part of the phase-space is untouched in the 
experimental analyses.
Let me focus on two examples.
All reported analyses aimed at discovering the mechanism 
of electro-weak symmetry breaking were limited to searches
for massive, point-like, scalar particle(s) which 
couple to fermions with a strength proportional to the fermion masses.
Consequently, only events with tagged heavy quarks 
were selected and searched for the  presence of narrow resonances.
To the same extent that this is unquestionably the optimal
standard-model-guided search strategy, leaving this strategy as the 
only one is difficult to accept, in particular 
from the point of view of  
an experimental, ''open-to-surprises" perspective.  
The mechanism by which fermions acquire masses in  the standard model,   
requiring  the introduction of the fermion-higgs coupling 
constants as arbitrary parameters of the model 
is, in my view, ''sufficiently  ugly"
not to constrain the curiosity of experimentalists.
As the second example, I chose searches in 
the high energy transfer (not necessarily 
correlated with the  high $Q^2$) frontier
of $ep$ scattering, 
a phase space region unique to HERA which 
was largely
untouched in {\it reported analyses} of the data.  
Leaving out this {\it difficult} fraction
of the phase space  has only  a weak impact on 
the derived exclusion limits
for the ''standard" beyond the standard model scenarios.
But, analysis of the data in this region 
cannot be omitted, if we  want to maximise the 
efficiency for detecting
unexpected novel phenomena rather than 
to establish the rejection limits
for the expected ones.

\subsection{Generic searches}
  
In view of the negative results of dedicated searches
several conference participants shared the opinion that   
{\it generic searches}
for new phenomena  should attract more attention.
  
The event selection criteria for such searches should, in my view, 
be optimised
to cover those phase space regions where the standard model
predictions can be made with sufficient precision 
to detect anomalies rather than
be optimised to
cover the phase space region where ''anomalies are expected"
and the ''standard model background" is easily controlled. 
Generic searches  diminish  the chances that
unexpected novel phenomena are overlooked if their manifestations in the
data do not follow one of the predefined scenarios and enlarge the
discovery potential if signatures of new phenomena are weak but present
in more than one initial and final state topology.

An example of generic searches was given in the talk of  
P.Savard \cite{Savard} representing the CDF and D$\O$ collaborations. In the 
analysis of the production of $W+b+ \bar b$, sensitive to 
single top production processes, 42 events were observed and 32 $\pm$ 5
events expected in the standard model. This is an important result    
for generic studies of multi-jet production accompanying
the propagation of gauge bosons in the electro-weak vacuum.    
Complementary and, in my view, indispensable data for such studies could
be collected by the LEP experiments in 
a dedicated 
large-luminosity run at the centre-of-mass energy 
below the threshold of the $W$-pair production - e.g. at 150 GeV. 

Last but not least,    
the generic analysis of high $E_T$ data at HERA
is very interesting  and the author
re-encourages the H1 and ZEUS experiments to consider such an 
approach. 


\section{High $E_T$ processes}
	 
Those of the  high $E_T$ processes, in which the total transverse
energy is shared between a small number of outgoing particles
and/or jets, are believed to be controlled by  perturbative QCD
to a high degree of precision. They define a "golden
domain" for testing  predictions of  perturbative QCD. 
The results presented at this conference covered production 
of high $E_T$ jets, photons, gauge bosons and heavy quarks.

\subsection{Jets}\label{subsec:jet} 
	
The preliminary CDF Run 1B inclusive jet cross-section data 
presented by C. Mesropian \cite{Mesropian} agree with the perturbative 
QCD predictions if one accepts  a large, 
but quite realistic (factor 2 
at $x=0.5$), uncertainty of the gluon density in the proton. 
The CDF jet $E_T$ spectra are in agreement with the D$\O$ data within the 
systematic errors quoted by both collaborations.
The weak $x_T$-dependence  of the 
ratio of the inclusive jet cross-sections at 1800 GeV and 630 GeV,
observed by both experiments at large $x_T = 2E_T/ \sqrt{s}$,
indicates that, if there is an excess of large $E_T$ jets,  it is 
related to an excess of partons 
carrying a large fraction of the proton momentum 
rather than 
to an increase of their cross-section in short distance processes.
It is noteworthy that the measured ratios are 
about 30 $\%$ lower than the perturbative QCD calculations.
Is the above  discrepancy 
related to the jet production mechanism? Does it reflect
the energy dependence of the energy flow 
uncorrelated with the production of a high $E_T$ jet? Or,
does it originate from  
the energy dependence of the transverse momentum of colliding 
partons? 
HERA experiments could, in my view, help not only in answering the above
questions
but also in constraining the large $x$ gluon distribution. 
The H1 and ZEUS electron tagging detectors, which 
measure  the energy of 
quasi-real photons colliding with protons, 
provide an  
experimental handle to study the 
energy dependent effects which could modify the 
jet transverse energy spectra. Similarly, extending 
the measurement of the jet $E_t$  spectrum up to 100 GeV 
(e.g. by accepting 
events in the full $y$ and $Q^2$ range) 
could constrain 
the gluon distribution in the $x$-region of interest for 
interpretation of the FNAL results.

C. Glasman \cite{Glasman}
representing the H1 and ZEUS collaborations  
discussed the ZEUS 
measurement of the inclusive jet $E_T$ spectrum
in photo-production. The data were shown to 
agree with the NLO QCD calculations
up to $E_T < 74$ GeV (for 
a sample of events constrained to the $ 0.2 < y < 0.85 , Q^2 < 4$ GeV$^2$
region). 
She also  reported the  H1 analysis of di-jet 
production in deep inelastic scattering.
The gluon distribution in the $0.01 < x < 0.1$ region 
derived in this analysis is  compatible with
the one determined by the scaling violation analysis of 
the $F_2$ structure function.

\subsection{Photons }\label{subsec:photons}

The compatibility of experimental results on 
inclusive high $E_T$ photon production and their interpretation
in perturbative QCD was discussed by M. Krawczyk \cite{Krawczyk},
 M.Werlen \cite{Werlen} and M.G. Strauss \cite{Strauss} 
who  represented the D$\O$ and CDF collaborations.
 
The NLO QCD calculation by Krawczyk and Zembrzuski 
provides a satisfactory description of the measured $E_T$
and $\eta$ spectra of isolated photons at HERA. 
The NLO QCD calculations by Vogelsang et al.  fail to describe
the FNAL measurements of 
the photon spectra
in the full range of the measured $E_t$ region - even if 
a wide range of variation of the renormalisation, factorisation
and fragmentation scales are allowed. The Gaussian smearing 
of partonic $k_T$ with $< k_T >= 3.5$ GeV improves the agreement.
This leads naturally to the question, 
'Could NLO QCD calculations
including re-summations 
 improve the agreement between the data and theoretical 
predictions?'. This question will remain open until such 
calculations are done.   
  
The recent critical study of the photon production 
processes by P. Aurenche
et al. \cite{dummy} and presented 
by M. Werlen demonstrates large discrepancies
between the data from various experiments.
In the low $x_T$ region the ISR  data are lower  by a factor 
of 2 to 4
than  the E706 data. The ISR data agree
with the NLO QCD calculations.  The E706 data, taken  
at a factor of $\approx 2$ smaller $\sqrt{s}$, 
require extra Gaussian smearing 
of partonic $k_T$ with $< k_T >= 1$ GeV. If the latter data are  
correct the question one could ask is, 'Is 
QCD able to predict the size and centre-of-mass energy evolution of
the effective $k_T$ smearing 
up to FNAL energies where $< k_T >= 3.5$ GeV?'.
In the large $x_T$ region the $x_T$-spectra differ by a factor of 2.
This is, at present, 
a direct measure of the uncertainty of the large 
$x$ gluon distribution.

\subsection{Gauge bosons} \label{subsec:bosons}

H. Melanson \cite{Melanson} presented the recent results on 
gauge boson production at FNAL.
Earlier FNAL measurements by  both 
the D$\O$ and the CDF collaborations, reported e.g. at Moriond-98, 
suggested a small excess of events in the $p_T > 60$ GeV region
with respect to the theoretical predictions 
(both for W and, to a lesser extent, for Z boson spectra).
This region is particularly 
interesting in view of the H1 events with large 
$p_T$ leptons and missing transverse energy
reported at this conference by C. Niebuhr \cite{Niebuhr}.
The new preliminary D$\O$ Run 1B results showed  at this conference
are in  perfect agreement with $O(\alpha_s^2)$, b-space
resummed calculations by Arnold and Kauffman \cite{dummy}. 
The improvement in the measurement precision
was reported to be due to ''a better 
understanding of hadronic jets with electron
signatures".

\subsection{Heavy Quarks }\label{subsec:heavy}	 

The CDF and D$\O$ measurements of the  b-quark 
$p_T$ spectra, presented
by  M. Baarmand \cite{Baarmand} show very good agreement.
The spectra were unfolded by the D$\O$ collaboration  from 
their di-muon and inclusive muon event samples and by the 
CDF collaboration
from their $J/\Psi$ and $\Psi (2S)$ data.
Given the perfect agreement between 
the resulting  spectra, the reported   
disagreement between these data and the predictions
of the NLO QCD (the data lay a factor 2.5 $\pm$ 0.4 above 
the theoretical prediction in the central region and
3.6 $\pm$ 0.8 in the fragmentation region) challenges
seriously our understanding of b-quark production 
processes in $p \bar p$ interactions 
within the framework of  the 
available perturbative QCD calculations.
The recent theoretical developments including the 
Variable Flavour Number scheme calculations by F. Olness
et al. \cite{dummy} and introduction of  a  harder $b \rightarrow B$ 
fragmentation function by Colangelo-Nason \cite{dummy}, does not provide 
a satisfactory explanation for 
the magnitude of this discrepancy.

A similar trend of the b-quark production cross-section 
being larger than expected was reported by M. Hayes
\cite{Hayes}
who summarised  the results of the HERA experiments.
The H1 and ZEUS  data were  compared to the LO Monte-Carlo
predictions. It would be interesting 
to see to which extent
including NLO corrections could  improve or worsen
the agreement between the data and the theory.
 
Last but not least I report the 
measurement of  the $p_T$ spectrum of t-quarks by the CDF
collaboration presented by P. Savard \cite{Savard}
and look forward to the prospects of measuring 
this spectrum in the forthcoming RUN II at FNAL.

\section{Rapidity gaps in $ep$ and $p \bar{p}$ collisions}

Events with rapidity gaps and  
high $p_T$ (high mass) particles
are grouped  at FNAL into  three classes.
Events with a gap in the $p$ or $\bar p$ fragmentation region 
are called {\it Hard Single Diffraction Events  (HSDE)}, those 
with two gaps in both the $p$ and   $\bar p$ fragmentation 
regions are called {\it Hard Double Pomeron Events (HDPE)}
and those with a single gap in the central region are
called {\it Hard Colour Singlet Events (HCSE)}.

M. Strauss \cite{Strauss} representing the CDF 
and the D$\O$ collaborations
showed the  final results on the fraction 
of HCSE events in a sample of events containing
high $E_T$ jets. The results on the ratio
of rapidity gap fractions at  630 GeV and 1800 GeV
of  3.4 $\pm$ 1.2 $\%$ (D$\O$) and 2.4 $\pm$ 0.9 $\%$ (CDF) 
are in a good agreement and can be interpreted in terms 
of decreasing survival probability of a gap 
with increasing energy.

 K. Goulianos \cite{Goulianos} 
summarised
the results of the CDF and D$\O$ experiments. He argued,  
on the basis of CDF results, that the fraction 
of events with rapidity gaps was to a good approximation 
independent of the nature and of the hardness scale  
of the process in which rapidity 
gaps were created. The list of these processes included:
W-production, di-jet production, b-quark production
and $J/ \psi$ production. The jet $E_T$ spectra
were reported to be compatible for double pomeron exchange,
hard single diffraction and non-diffractive
classes of events. 
In addition 
the relative ''HDPE"/''HSDE" and
''HSDE"/''ALL" fractions were found to be the same within 
experimental errors. 

The processes producing leading  baryons
and/or rapidity gaps in $ep$ collisions
at HERA were discussed by M. Martinez \cite{Martinez}
in a formalism based on the  
$F_2^{D(3)}$, $F_2^{LN(3)}$ and $F_2^{LP(3)}$
structure functions.
Leaving aside the controversial formalism used in the 
analysis,
the data clearly demonstrated 
that the 
mechanism producing rapidity gaps, leading  protons or neutrons 
is decoupled  
from the process of deep inelastic collision  of 
the projectile  
electron with  one of the charged soft partons. 
This statement is more general than the  the Regge-motivated 
factorisations of $F_2^{D(3)}$, $F_2^{LN(3)}$ and $F_2^{LP(3)}$
and, in particular,  does not preclude a specific ordering of 
reggeon formation and deep inelastic electron-reggeon collision.

In my view, both the HERA and Tevatron
data  provide increasing evidence that 
in a large variety of hard diffractive processes   
the same point-like   
structure of the vacuum excitation
is probed. 
This structure appears to be universal i.e.   
independent of whether a particular 
de-excitation  mode      
produces  rapidity 
gaps, leading neutrons or protons.
Such universality can be extended to other
modes provided that they  are not correlated with the 
azimuthal angle of the hard scattering plane.
It is my hope that the experimental studies of
such low-frequency lepton-induced de-excitation modes  
of the QCD media of 
variable  colour-field strength (generated
by nuclei of variable atomic number) 
will give in the future an important  
boost in our understanding of QCD
at large distance scales.

 \section{Structure functions}
\subsection{Nucleon. }\label{subsec:nucl}

The measurements of the double differential cross-section 
$d \sigma/dxdQ^2$ in electron-proton scattering cover
a very impressive range of the $(x,Q^2)$ domain. M. Costa \cite{Costa} 
presented the results from ZEUS and H1 
experiments which extend the measurements to 
the low $Q^2 \geq 0.065$ GeV$^2$ region.
After several years of continuous progress in 
improving the accuracy of the data   
the point-like charge structure of the proton 
is mapped with high precision by the HERA experiments.
On the other hand, unfolding the    
relative contributions of the photo-absorption cross-section 
for  longitudinally and transversely
polarised photons and 
interpreting the resulting structure functions in 
the framework of QCD, in particular in the low $x$ domain, 
are not free of ambiguities. 
 
The method used by the H1 collaboration  
to determine the contribution of 
longitudinal photons to the measured cross-section 
is, in my view, controversial and
cannot replace  
the conventional Rosenbluth separation method 
which requires varying the HERA
beam energy (-ies). 
The reported success in applying the 
DGLAP equation to describe  the $Q^2$-evolution
of the measured $F_2$ structure function 
down to  $Q^2$ values as low 
as 1 GeV$^2$ can hardly
be considered as equivalent to confirming its underlying 
partonic dynamics. 
The DGLAP fits of the 
deep inelastic scattering 
data have
substantial freedom in absorbing 
a wide range of  ''non-DGLAP" $Q^2$-shapes  of $F_2$ by the fitted
$x_{Bj}$-shape of the gluon distribution at fixed $Q_o^2$ scale
- in particular in the low $x_{Bj}$ region where the $Q^2$ span of
measured $F_2$ values is small.
The $x_{Bj}$-shape  of the gluon distribution at 
$Q^2 = 1$ GeV$^2$, showed by 
M. Costa \cite{Costa}
is in my opinion 
sufficiently ''unphysical" to  
indicate  that non-perturbative scaling violation effects
and/or non-linear QCD effects e.g. those  
discussed at his conference by R. Venugopalan \cite{Raju} are 
present in the data. 

Another result  which, in my view, 
raises serious doubt  on the applicability
of the  DGLAP equations 
for HERA data below $Q^2 = 10$ GeV$^2$
was presented  in a talk by G. Marchesini's \cite{Marchesini}  
on new developments in power corrections.
One of several issues discussed in this talk  
was the Breit-frame analysis of 
the fragmentation spectra of the current quark in 
deep inelastic scattering.
The  power corrections 
to the quark fragmentation function
modify, in this kinematical region,  the NLO QCD predicted values by 
factors up to 20. Leaving aside explaining 
the source of such large corrections
\footnote{I devoted a large fraction of my
          talk at the recent RIKEN Workshop on "Hard 
	  Partons in High Energy Nuclear Collisions"
	  to this issue.}
their impact on the $F_2$ values determined by methods based
on the measurement  and  QCD-Monte-Carlo simulation
of the hadronic system must be
large. Why are there no traces of these large 
corrections in the comparison of $F_2$ values
determined using electron and hadronic methods?
 
I would like  to summarise the discussion of the HERA 
deep inelastic scattering results by two 
conclusions. 
The first is that 
one should use with  caution the existing parametrisations
of the gluon distribution in the very low x region.
The second, optimistic one, is that we may have 
already entered a new QCD regime of saturated
partonic densities. 
 
The QCD analyses of 
deep inelastic scattering data sets
leave a substantial uncertainty 
in the derived gluon distribution not only at low $x$ but,
as discussed by A. Ball \cite{Ball}, as well  at large $x \geq 0.3$.
In this region non-perturbative effects are absorbed,
by common conventions 
rather than by common understanding of their size,  
into  higher twist and target mass corrections. 
The {\it a priori} unknown size of  higher twist corrections
\footnote{ An ambitious program aiming at  
 understanding the magnitude of such 
corrections was presented by G. Marchesini \cite{Marchesini}.}  
give rise to an uncertainty in assigning a fraction of 
observed $Q^2$ evolution of the cross-sections 
to perturbative, gluon mediated processes. The
relative importance of perturbative and non-perturbative
contributions to structure functions were discussed by S. Alekhin
\cite{Alekhin} who presented 
 a new analysis 
of the fixed target deep inelastic scattering data. 
As an example, the leading twist contribution 
to $R= \sigma_L / \sigma_T$ at $Q^2 = 2$ GeV$^2$ and 
$x_{Bj}=0.5$
is below 10 $\%$ of the measured value. It 
illustrates how loosely 
the $x$-distribution of the gluon is constrained by 
a rather precise R measurement.   
     
Global QCD fits of a large variety of hard process data,
provide  the input parton distribution functions
for commonly used Monte-Carlo generators.
In addition they cross-check the consistency 
of data sets which cannot be directly compared.
A. Ball  \cite{Ball} reported recent 
news coming from  this field of activity.
He expressed his vision of the bright future 
by  quoting D. Kosower's: 
"...end of the tyranny of the global fitters"  
call at the recent La Thuille conference
and by explaining a new fitting method which,
in my view,  
could create "{\it a  brave new  world}
of data sets" for fitters.
The new  method re-addresses a very important and 
very difficult problem: how to   
include correlated systematic 
experimental errors in the fitting procedure 
in order to properly assess the size of systematic errors
of the resulting partonic distribution functions?
This has been  
a controversial subject over the last 15 years 
and will most likely stay controversial.
Instead of addressing technical aspects
of the old and new method let me 
present the following {\it phenomenological} observation  
which shifts the centre-of-gravity of the problem.
Contrary to {\it old-fashioned expectations},     
the ''quality" of several  
deep inelastic data sets which 
determine the  partonic distribution functions, 
appears to be proportional to the number and the size 
of experimental corrections determined with the 
help of QCD Monte-Carlo simulations,
which leave no traces in the error matrices
of the data points.

\subsection{Polarised nucleon }\label{subsec:pol}

The preliminary results of the E155 experiment were 
reported by R. Erbacher \cite{Erbacher}.
They included  measurements of the $g_1$ and the $g_2$ structure 
functions of the proton and of the deuteron 
and the results of the NLO QCD fits to the data.
 The Bjorken sum rule 
is confirmed by the high precision data. The measured 
 $ \Gamma_1^{p-n} = 0.172^{+0.005~+0.008}_{-0.003~-0.007}$
 agrees with the predicted value of
  $ \Gamma_1^{p-n} = 0.182 \pm 0.005$.
  
An attempt to directly determine the contribution of gluons to the 
spin of the nucleon has been made by the Hermes collaboration 
and reported by J. Martin
 \cite{Martin}. Their value of 
$\Delta G/G = 0.41 \pm 0.18 (stat) \pm 0.03 (syst)$
 however relies heavily on the modelling of hadron production processes     
in {\it ep} scattering down to $p_T = 1.5$  GeV. The
PYTHIA Monte-Carlo was used to determine  
the relative contributions of photon-gluon
fusion, QCD-Compton and soft VDM processes to the observed 
$p_T$ spectra of charged hadrons.

\subsection{Photon and Virtual Photon}\label{subsec:phot}

Measurements of the leptonic photon structure function 
$F_{2,QED}^{\gamma }$ by the OPAL and L3 collaborations were
presented by M. Chamizo \cite{Chamizo}. 
This structure function, contrary to 
the hadronic photon structure function,
can be derived within QED. It is interesting to note that,
the average virtualities  of the probed photon
of $<P^2>= 0.034$ GeV$^2$ (L3) and $<P^2>= 0.05$ GeV$^2$ (OPAL), 
required  to achieve  
good agreement between the  theoretical 
prediction and the data are both compatible with $4*m_{\mu }^2 $
and that the
contribution of longitudinal photons can not be neglected.

M. Chamizo \cite{Chamizo}
showed discrepancies 
between the theoretical predictions and the measurements
of the hadronic photon structure function $F_{2,QCD}^{\gamma }$
by  LEP experiments. Discrepancies were also reported 
by T. McMahon \cite{McMahon} who presented measurements 
of the di-jet cross-section in photo-production
at HERA. 
It will be interesting to see if 
they could  be 
absorbed into a new set of partonic densities of the photon
which would   
describe both the HERA and the LEP  photon-structure-dependent 
observables.

Production of jets in deep inelastic ep scattering, 
in particular in processes  in which a jet 
is  emitted at a  large 
$\Delta \eta = \eta_ {jet} - \eta _{QPM}$
have attracted  considerable attention at HERA.
These processes have been expected to  provide evidence
for  so called ''BFKL-dynamics" - one of buzz-words
at HERA.  
M. Swart \cite{Swart} representing the  H1 and ZEUS collaborations 
showed the comparison  of the jet spectra with the recent 
NLO QCD calculations by B. Poetter 
\cite{dummy}. Good agreement was found
at the price of introducing a non-perturbative
structure of the virtual photon.  
T. McMahon \cite{McMahon}
used the H1 di-jet production data
to determine the effective partonic densities of the virtual photon
in  the ''Leading Order Single Effective
Subprocess Approximation". 
One of several  possible conclusions from the above studies, 
for those who find it difficult to 
accept absorbing our lack of understanding into a new set 
of structure functions,   
is that  HERA processes with two 
hard scales $Q^2$ and $p_T^2$,   keep on challenging perturbative QCD.

\subsection{Pomeron }\label{subsec:pom}

The pomeron structure function $F_2^{D(2)}$
and its partonic interpretation 
based  on the fits to the H1 rapidity gap data
were discussed by  M. Martinez \cite{Martinez}.
The pomeron is mostly glue:
at $Q^2= 4.5$ GeV$^2$ 90 $\%$ of the pomeron momentum
is carried by gluons. This fraction is comparable 
to the fraction of the proton momentum carried by gluons
in the low $x_{Bj}$ region where the bulk of
rapidity gap events is observed. 
  
How universal is the measured pomeron structure?
Can one use partonic densities of ''HERA pomerons"
in other diffractive processes? Do we need 
to introduce  pomeron partonic structure to 
describe diffractive processes at all?
There are two reasons why I ask these questions.
The first boils down to  an  aesthetic ''Occam-razor" argument,   
which forbids  adding the pomeron 
to the list of hadronic objects with distinct 
partonic structures before verifying if such structure cannot 
be simply derived from  those already included in the list.
To my best knowledge no compelling experimental evidence
has been presented so far by the HERA 
experiments that the pomeron structure 
cannot be explained by the proton structure 
at $x_{Bj}= \beta x_{pom}$ and fixed $Q^2$.   
The second one  is driven by 
an experimental observation presented at this 
conference.
It is natural to expect 
that the partonic structure of pomerons
produced at HERA  is sufficiently universal to be 
applicable to rapidity gap processes observed at FNAL.
The 
preliminary 
analysis of the data of the CDF collaboration on jet production
in rapidity gap events which was 
reported by K. Gulianos \cite{Goulianos}
indicated that this might not be the case.
The comparison of the data with the predictions
based on partonic distributions of the ''H1-pomerons"
showed large  {\it $\beta $-dependent}  discrepancies
suggesting  a lack of 
such universality.
Even if these  discrepancies can eventually  be 
be explained by the subleading reggeon trajectories
such  phenomenology becomes
hardly predictive and of rather limited merit.

\section{Strong interactions in nuclear media}

P.Seyboth's \cite{Seyboth} statement summarises best 
the experimental results on 
heavy ion collisions. He said    
that "the SPS data  were 
{\it probably compatible} with a QGP phase transition but 
the efforts of experimentalists to provide confirmation 
of a clear threshold in energy or nuclear size and the 
efforts of theorists 
on alternative interpretations of the observations
have to be  pursued".

L.Kluberg \cite{Kluberg} presented 
several consistency checks of  
the data of the NA50 experiment. 
One of the most important was the analysis of the 
target length dependence of the ratio of $J/ \Psi$ to Drell-Yan
cross-sections. Suppression of $J/ \Psi$ production in Pb-Pb
central collisions (with respect to extrapolations from 
lighter nuclei) was demonstrated to be robust against several
experimental checks. Whether it can be considered as a 
signature of the phase transition remains an open question.
I would like to remind the reader at this point that  
the impact parameter dependence
of the ratio of quark and gluon distributions in large nucleus,
has never been measured. This ratio determines
the relative strength of Drell-Yan and $J/ \Psi$ production.
Its extrapolation  
to central Pb-Pb collisions,
using only the impact-parameter-integrated
A-dependent quark distributions, is a subject of uncertainty which,
in my view, 
 cannot be precisely assessed at present.

The CERES results on $e^+e^-$ production in p-A and A-A collisions
have been reported by T. Wienhold \cite{Wienhold}. Their theoretical 
analysis was presented  by H. Hansson \cite{Hannson}.
The enhancement in the $e^+e^-$ mass spectrum 
with respect to pp-scaled sources in the region 
of $0.25 < m_{ee} < 0.7$ GeV of $3.9 \pm 0.9 (stat) \pm 0.9 (syst)$
observed in the 1995 data is confirmed by the 1996 data 
(the reported excess is $2.6 \pm 0.5 (stat) \pm 0.5 (syst)$).
The model which was used in the extrapolation 
agrees very well with $e^+e^-$ mass  spectra observed
in p-Be and p-Au collisions. 
The particle ratios were taken from a thermal 
model fitted to the measured ratios in Pb-Pb collisions.    
The enhancement is most pronounced at low $p_T$
of the pair: $p_T < 0.5 $ GeV.

In Quark Gluon Plasma (QGP) one expects  the relative multiplicity of 
strange versus non-strange particles to increase with 
increasing  
strangeness content of produced baryons. The WA-97 data presented
by P. Norman \cite{Norman} exhibited  clearly such a behaviour. However,
this explanation of the data is not  unique.
The Dual Parton Model analysis
presented by C. Salgado \cite{Salgado}
explains the hyperon yields 
by standard non-QGP processes. In addition, 
 A. Rybicki \cite{Rybicki} showed that the strangeness
enhancement was observed already in the central p-Pb
collision NA49 data in the fragmentation region
in contradiction to the VENUS Monte-Carlo predictions.

One of the most interesting measurements presented at the 
XXXIV Moriond was the first deuteron , anti-deuteron and triton
coalescence results of the NA44 collaboration
discussed  by J.J. Gaardhoje \cite{Gaardhoje}.
Low energy anti-dueterons can be very efficiently identified
by NA44 by  using the standard time-of-flight method 
and, in addition, 
by measuring the excess of energy deposited in the calorimeter due
to the extra energy released in the $\bar d$ annihilation.
In order to produce deuterons (anti-deuterons)  the 
recombination of nucleons (anti-nucleons) must take  place 
at the very late stage of particle production phases so as 
 not to disrupt such  loosely bound nucleon system.
The freeze-out  radius
was shown to be 
compatible with the interferometric  measurement
of the size of the pion source.

The nucleus can be used as a femto-vertex detector in studies
of the space-time structure of strong interactions,
in particular in processes with point-like leptonic
probes.
The HERMES eA data on $ \rho$ production presented by T. Shin
\cite{Shin} showed  evidence
for the  variation of the nuclear transparency 
ratio with the coherence length.

\section{Concluding remarks} 

Investigation of ''colourful" strong interaction 
phenomena is sufficiently challenging to give us a lot of fun.
While the experimental studies of electro-weak phenomena put an emphasis  
on the detector design and performance, 
optimised for the highest possible accuracy
of {\it the  predefined measurements}, 
the research in  the strong interaction domain,
in particular in its most challenging {\it confinement 
and QCD-vacuum sectors},
need, first of all, a fresh, curiosity-driven, and open-to-surprises
look at the data. 

The lack of theoretical guidelines in defining ''appropriate
variables" 
which could link short-distance  and long-distance,
strong interaction phenomena makes this task 
hard. But, at the same time it gives us, 
experimentalists, a challenging 
chance to use our imagination rather than our fingers 
attached to the workstation keyboards while scrutinising
the agreement between the corresponding ''Ntuple variables''
of the  Monte-Carlo  and the data.

Moriond meetings stimulate the  atmosphere, in which admitting 
the lack of understanding rather than hiding it under the 
cover of self-assuring consensus is of value. They give us 
the necessary enthusiasm to face the challenges. 
Moreover,  they  create hope  
that the forthcoming studies of strong interaction
phenomena will 
keep on bringing new
interesting  
and unexpected results for the next millennium Rencontres.     
 
\section*{Acknowledgments}
I would like to thank  J. Chwastowski, R. Devenish, B. Klima,  G. Myatt and
D. Waters for critical reading of the manuscript.

\section*{References\footnote{All 
                         listed below contributions will be published in  
                         the Proceedings of the XXXIVth 
                         Rencontre de Moriond, edited by 
			 Jean Tr\^an Thanh V\^an, Les Arcs,
			 March 1999.}}

\end{document}